\documentclass[%
 aip,
 sd,%
 amsmath,amssymb,
 reprint,superscriptaddress,%
]{revtex4-1}

\usepackage{graphicx}
\usepackage{dcolumn}
\usepackage{bm}

\begin{document}


\title{Design of geometric phase measurement in EAST Tokamak}
\author{T. Lan}
\affiliation{Department of Modern Physics and School of Nuclear Science and Technology, University of Science and Technology of China, Hefei, Anhui 230026, China}
\affiliation{Institute of Plasma Physics, Chinese Academy of Sciences, Hefei, Anhui 230031, China}%
\author{H. Q. Liu}
 \email{hqliu@ipp.ac.cn.}
\affiliation{Institute of Plasma Physics, Chinese Academy of Sciences, Hefei, Anhui 230031, China}%
\author{J. Liu}
\affiliation{Department of Modern Physics and School of Nuclear Science and Technology, University of Science and Technology of China, Hefei, Anhui 230026, China}%
\author{Y. X. Jie}
\affiliation{Institute of Plasma Physics, Chinese Academy of Sciences, Hefei, Anhui 230031, China}%
\author{Y. L. Wang}
\affiliation{Department of Modern Physics and School of Nuclear Science and Technology, University of Science and Technology of China, Hefei, Anhui 230026, China}%
\author{X. Gao}
\affiliation{Institute of Plasma Physics, Chinese Academy of Sciences, Hefei, Anhui 230031, China}%
\author{H. Qin}
\affiliation{Department of Modern Physics and School of Nuclear Science and Technology, University of Science and Technology of China, Hefei, Anhui 230026, China}%
\affiliation{Plasma Physics Laboratory, Princeton University, Princeton, NJ 08543, USA}%
\date{\today}
\begin{abstract}
The optimum scheme for geometric phase measurement in EAST Tokamak is proposed in this paper. The theoretical values of geometric phase for the probe beams of EAST Polarimeter-Interferometer (POINT) system are calculated by path integration in parameter space. Meanwhile, the influences of some controllable parameters on geometric phase are evaluated. The feasibility and challenge of distinguishing geometric effect in the POINT signal are also assessed in detail.\\
\end{abstract}
\maketitle
\section{\label{section 1}Introduction}
\indent The geometric phase, as a fundamental phenomenon, has attracted a great deal of research interests in different subfields of physics. The geometric phase in quantum systems, dubbed Berry phase, was systematically introduced for the first time by Berry in 1984.\cite{Berry1984Quantal}After that, its deep geometric origin and wide existence in classical mechanics were discovered.\cite{Hannay1985Angle}A variety of experiments has been carried out to verify the existence of geometric phase in different systems repeatedly.\cite{Tomita1986Observation}Meanwhile, the geometric phase has been applied to the development of advanced material and quantum information techniques. In plasma physics, some theoretical investigations on geometric phase have recently been achieved, such as the geometric phases associated with the gyro-motion of charged particles and the electromagnetic waves in non-uniform magnetized plasmas.\cite{Littlejohn1988Phase,Bhattacharjee1992,Liu2011Geometric,Liu2012Geometric}However, there's still no direct observation of geometry phase in plasma experiments.\\
\indent On the one hand, the experimental observation of geometric phase in plasma systems is an essential verification of the geometric phase theory by a new experimental technique, which has important value in methodology. On the other hand, the measurement of geometric phase confirms geometric effect as a new system error in the existing diagnostics, which benefits a more complete understanding of the error sources in polarimetry system. Because of its geometric origin, the geometric phase has totally different properties from the dynamical phase, which arises directly from the dynamical process of the system. The different properties between the geometric phase and dynamical phase can be used to distinguish them in experiments. The geometric phase in Faraday rotation angle for linearly polarized electromagnetic waves propagating in non-uniform magnetized plasmas is a good candidate for the first identification of geometric phase in plasma because of the development of the polarimeter-interferometer hardware system.\cite{Liu2012Geometric}The popular three-wave polarimeter-interferometer technique has been widely used in the polarimeter-interferometer systems of several main magnetic confinement fusion devices, such as MST \cite{Ding2010Upgrade}, RTP\cite{Rommers1997The}, J-TEXT\cite{Chen2012First}, EAST\cite{Liu2013Design,Liu2014Faraday,Zou2014Optical}, and the next-generation tokamak ITER\cite{Yamaguchi2008Optimization,Imazawa2011Thermal,Donne2004Poloidal,Kondoh2004Toroidal}.\\
\indent The Faraday effect is caused by different phase velocities of two eigen waves, i.e., the right-handed (R) and left-handed (L) waves. In non-uniform magnetized plasmas, the two eigen waves, as well as the linearly polarized wave, undergo anholonomic processes and hence include geometric phases. The geometric phase in Faraday rotation angle depends only on the wave trajectory instead of the specific dispersion relations. Since the geometric phase cannot be inferred from the wave equation directly, it is easily overlooked. Because the conventional three-wave polarimeter-interferometer technique only takes into account the dynamical phase, which is calculated using the dispersion relations, the geometric effect resulting from the deviation of wave trajectories and non-uniformity of the magnetized plasmas is ignored. The missing geometric effect turns out an error source in the existing diagnostic systems. The calculation of the error range of the polarimetry system caused by ignoring the geometric phase under typical high operation parameters of EAST tokamak contributes to gain an insight into the influence of geometric effect on diagnosis accuracy of polarimetry system.\\
\indent In this paper, the geometric phase in EAST Polarimeter-Interferometer (POINT) system is evaluated using typical high operation parameters of EAST tokamak. The feasibility of distinguishing geometric phase in the signals of POINT system is analyzed in detail. To estimate the realistic geometric phase in POINT system, the beam trajectories in non-uniform magnetized plasma is calculated using the ray-tracing simulations. The geometric phase can be obtained by computing the integrals along the wave trajectory in the parameter space. According to the simulation results, the geometric phases in the signals of present EAST POINT system ranges from $1\times10^{-3}$ to $1\times10^{-2}$ degree, which is much smaller than the resolution of POINT system, i.e., 0.1 degree. It is found that the geometric phase can be magnified by adjusting the toroidal angle of the incident beam. Several schemes are proposed for the measurement of the geometric phase in POINT system by amplifying the geometric phase and enhancing the diagnostic resolution.\\
\indent The paper is organized as follows. In Sec.~\ref{section 2}, the theory for data processing in three-wave polarimeter-interferometer technique is improved by considering the existence of the geometric phase. In Sec.~\ref{section 3}, the realistic geometric phase in the double-pass system is analyzed. And the influence of retro reflectors (RR) on the magnitude of the geometric phase is also discussed. In Sec.~\ref{section 4}, the range of the geometric phase is evaluated using the single-pass setup of POINT system. The feasibility and challenge of distinguishing geometric phase from the signals of POINT system are studied. In Sec.~\ref{section 5}, we give a brief conclusion and propose the optimum scheme of measuring the geometric phase using POINT system.\\

\section{\label{section 2}Geometric phase effect in three-wave polarimeter-interferometer technique}
\indent The three-wave polarimeter-interferometer technique, which was originally proposed by J. H. Rommers and J. Howard, obtains phase shift and Faraday rotation angle with two counter-rotating circularly polarized waves and one linear polarized wave.\cite{Rommers1996A} According to this method, line-integrated density and line-integrated poloidal magnetic field can be obtained using the following equations\\
\begin{eqnarray}
\label{eq:1}
\frac{\phi_L+\phi_R}{2} &=& C_1 \lambda \int{n_e} dL,
\end{eqnarray}
\begin{eqnarray}
\label{eq:2}
\frac{\phi_L-\phi_R}{2} &=& C_2 \lambda^{2}\int{n_eB_{\parallel}}dL,
\end{eqnarray}

\indent where $\phi_L$ and $\phi_R$ denote the phase shift for L- and R- waves respectively, $B_{\parallel}$ is the component of the equilibrium magnetic field parallel to the wave trajectory, $n_e$ is the electron density of plasmas, and $\lambda$ is the wavelength of the probe beam. In inhomogeneous magnetized plasmas, the deviation of the propagation direction of electromagnetic waves results in the extra geometric phase, which is not included in Eqs.~\ref{eq:1} and \ref{eq:2}. The type I geometric phases for left-handed and right-handed circularly polarized waves are determined by \cite{Liu2012Geometric}
\begin{eqnarray}
\label{eq:3}
\theta_g ^{L,R}=\pm \int_{P} d\bm{e_2 \cdot e_1}.
\end{eqnarray}
\indent Here $\bm{e_1}$ and $\bm{e_2}$ are two orthogonal unit vectors within the polarization plane, and the superscript L and R denote the left-handed and right-handed polarized waves, defined with respect to the wave vector. Since Eqs.~\ref{eq:1} and \ref{eq:2} are not the complete solutions of the wave propagation in the non-uniform magnetized plasmas, we can revise them by counting in the geometric phase terms as\\
\begin{eqnarray}
\label{eq:4}
\frac{\phi_L+\phi_R}{2} &=& C_1 \lambda \int{n_e} dL+\frac{\theta_g^{L}+\theta_g^{R}}{2},
\end{eqnarray}

\begin{eqnarray}
\label{eq:5}
\frac{\phi_L-\phi_R}{2} &=& C_2 \lambda^{2}\int{n_eB_{\parallel}}dL+\frac{\theta_g^{L}-\theta_g^{R}}{2},
\end{eqnarray}

\indent where $\theta_g^{L}$ and $\theta_g^{R}$ are the type I geometric phases corresponding to L- and R- waves respectively.\\
\indent The trajectories of L- and R- waves in the three-wave polarimeter-interferometer diagnostic are approximatively overlapped. Because the geometric phases for collinear counter-rotating circularly polarized waves are opposite, see Eq.~\ref{eq:3}, the geometric phase enters the Faraday rotation signal but vanishes in the phase shift signal in three-wave polarimeter-interferometer diagnostics.\\
\indent At the same time, the sudden change of propagation direction caused by optical systems can also lead to discrete geometric phases.\cite{Haldane1987Comment}This source of geometric phase is almost changeless once the optical arrangement is fixed. So the geometric phases arising from the optical systems could be readily eliminated in signal processing. And we will not focus on this part of geometric phases in this paper.\\
\section{\label{section 3}The influence of retro reflectors on geometric phase: Double-pass propagation}
\begin{figure}[htbp]
\includegraphics[width=.48\textwidth]{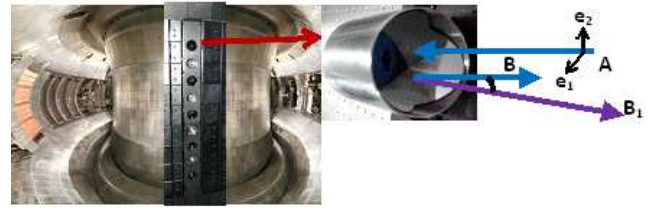}
\caption{The picture and illustration of retro reflectors of POINT system in the EAST tokamak.}
\label{fig:1}
\end{figure}

\indent In POINT system, molybdenic retro reflectors are installed on the inner vessel wall separately to guarantee that each probe beam returns to the detector precisely, see Fig.~\ref{fig:1}.\cite{Lan2015Study}The incident beam A and the reflected beam B propagate in approximatively opposite directions after triple reflections.\\
\\
\indent For incident beam A, L- and R- waves are defined as $\bm{e_1} \mp i\bm{e_2}$ with respect to the wave vector. Here $\bm{e_1}$ and $\bm{e_2}$ are orthogonal unit vectors within the polarization plane, satisfying $\bm{e_1} \times \bm{e_2}=\bm{e_{k_A}}$. According to the perfect conductor boundary condition, the polarizations of the reflected beam B and the incident beam A are the same in the laboratory frame. The electric components of the linearly polarized incident beam A and the corresponding reflected beam B can be decomposed into the eigen waves as\\
\begin{subequations}
\label{eq:6}
\begin{eqnarray}
&\bm{E_A}\!=\!E \left[ (\bm{e_1}\!-\!i\bm{e_2})e^{i\phi_A (\omega_1)}\!+\!(\bm{e_1}\!+\!i\bm{e_2})e^{i\phi_A (\omega_2)}\right]\!,
\label{subeq6:1}
\end{eqnarray}
\begin{eqnarray}
&\bm{E_B}\!=\!E \left[ (\bm{\!-\!e_1}\!+\!i\bm{e_2})e^{i\phi_B (\omega_1)}\!-\!(\bm{e_1}\!+\!i\bm{e_2})e^{i\phi_B (\omega_2)}\right]\!.
\label{subeq6:2}
\end{eqnarray}
\end{subequations}
\indent The Faraday rotation angles of beams A and B are respectively\\
\begin{eqnarray}
\label{eq:7}
\left\{
\begin{aligned}
\psi_A &=\frac {\phi_A (\omega_1)-\phi_A (\omega_2)}{2}, \\
\psi_B &=\frac {\phi_B (\omega_1)-\phi_B (\omega_2)}{2}+\pi. \\
\end{aligned}
\right.
\end{eqnarray}
\indent Note that the L- and R- waves in dispersion relations is defined with respect to the background magnetic field, which is different from the conventional definition of the left-handed and right-handed waves in optics, see Eqs.~\ref{eq:3} and Eqs.\ref{eq:10}. The left-handed polarized wave is L-wave if the beam propagates along the background magnetic field, and R-wave if it propagates against the magnetic field. Because of the difference definitions, the sign of dynamical phases for L- and R- waves keeps the same after reflection. But the sign of geometric phases become opposite after reflection. The phase shift of beams A and B can be expressed as\\
\begin{eqnarray}
\label{eq:8}
\left\{
\begin{aligned}
\phi_A (\omega_1) &=\theta_{d\!-\!A}^{L}(\omega_1)+\theta_{g\!-\!A}^{L}(\omega_1), \\
\phi_B (\omega_1) &=\theta_{d\!-\!B}^{L}(\omega_1)+\theta_{g\!-\!B}^{R}(\omega_1), \\
\phi_A (\omega_2) &=\theta_{d\!-\!A}^{R}(\omega_2)+\theta_{g\!-\!A}^{R}(\omega_2), \\
\phi_B (\omega_2) &=\theta_{d\!-\!B}^{R}(\omega_2)+\theta_{g\!-\!B}^{L}(\omega_2). \\
\end{aligned}
\right.
\end{eqnarray}
\indent Meanwhile, the total Faraday rotation angle is\\
\begin{eqnarray}
\label{eq:9}
\begin{split}
&\psi \!= \!\psi_A+\psi_B \\
&\!= \!\frac{\left[\theta_{d\!-\!A}^{L}(\omega_1)\!+\!\theta_{d\!-\!B}^{L}(\omega_1) \right]\!-\!\left[\theta_{d\!-\!A}^{R}(\omega_2)\!+\!\theta_{d\!-\!B}^{R}(\omega_2) \right]}{2} \\
&\!+\!\frac{\left[\theta_{g\!-\!A}^{L}(\omega_1)\!+\!\theta_{g\!-\!B}^{R}(\omega_1) \right]\!-\!\left[\theta_{g\!-\!A}^{R}(\omega_2)\!+\!\theta_{g\!-\!B}^{L}(\omega_2) \right]}{2}+\pi.
\end{split}
\end{eqnarray}
\indent If the wave trajectories of beams A and B overlap exactly, the geometric phase will cancel out in double-pass propagations. However, in practice the RRs alter the reflected beam trajectories in two aspects. On the one hand, the processing error of RRs results in non-parallelity between the incident and reflected beams.
On the other hand, the deviation between the incident beam and the center of RRs leads to the offset between the incident beam and reflected beam at the output aperture of RRs, see the purple arrow in Fig.~\ref{fig:1}. Considering the incident beam and reflected beam are symmetrical about the center of RRs after triple reflections \cite{Arnold1979Method}, the offset between the incident beam and reflected beam is twice the offset between the incident beam and the center of RRs at the output aperture of RRs. Under typical operation parameters of EAST tokamak, the offset between the incident beam and the center of RRs at the output aperture of RRs is of the order of centimeters, and the offset between the incident beam and reflected beam at the output aperture of RRs is about several centimeters. Therefore, the geometric phase would evidently participate in the Faraday rotation signal of POINT system even in the double-pass three-wave polarimeter-interferometer diagnostics.\\
\indent On the other hand, the present POINT system can also operate in single-pass setup, by reversing the polarization direction of reflected beam. This optic arrangement of POINT system can be achieved by installing some optical elements in the inner vessel wall. In the single-pass setup, the cancelling of the geometric phase due to assumed symmetries can be avoided. The range of the geometric phases can then be properly estimated. In Sec. IV, we will evaluate the geometric phases in POINT system using the single-pass setup.\\

\section{\label{section 4}Estimations of the geometric phases in EAST POINT system}
\indent For collisionless cold plasma, the dispersion relations for R- and L- waves are\\
\begin{eqnarray}
\label{eq:10}
\bm{N^2}=1-\sum_\alpha \frac{\omega_{p\alpha}^2}{\omega(\omega\mp\omega_{c\alpha})}.
\end{eqnarray}
\indent Here $\omega_{p\alpha}$ and $\omega_{c\alpha}$ denote the plasma frequency and gyro-frequency for the $\alpha$-th particle species respectively. Given cylindrical approximation, the equilibrium magnetic field can be expressed as\\
\begin{eqnarray}
\label{eq:11}
\bm{B}=\frac{B_0R_0}{R}\bm{e_\varphi}+\frac{rB_t}{qR}\bm{e_\theta},
\end{eqnarray}
\indent where $\bm{e_\varphi}$ and $\bm{e_\theta}$ are unit vectors along toroidal and poloidal directions respectively. If the characteristic length-scale and time-scale of the non-uniform plasma are much larger than wave-length and period of the incident wave, the following Ray tracing equations are available for the description of wave trajectories,\\
\begin{eqnarray}
\label{eq:12}
\left\{
\begin{aligned}
\dot{\bm{K}}&=\frac{\partial D}{\partial \bm{L}}/\frac{\partial D}{\partial \omega}, \\
\dot{\bm{L}}&=-\frac{\partial D}{\partial \bm{K}}/\frac{\partial D}{\partial \omega},
\end{aligned}
\right.
\end{eqnarray}
\indent where the dielectric tensor $D\!=\!k^2\!-\!\omega^2\bm{N^2}/{c^2}$ reflects the local property of plasmas. The profiles of electron density and safety factor are assumed as\\
\begin{eqnarray}
\label{eq:13}
q=C_3r^{N_q}+C_4,\\
\label{eq:14}
n_e=C_5r^{N_n}+C_6.
\end{eqnarray}
\indent Combining Eqs.~\ref{eq:10} to Eqs.~\ref{eq:14}, the trajectories of probe beams in magnetized plasmas can be numerically calculated. The type I geometric phase $\theta_g$ can hence be achieved by path integral in the parameter space.\\
\indent In the case that parameter distributions are symmetrical, most part of geometric phase in double-pass signal would be eliminated. If the distributions are asymmetric because of some local asymmetries, the geometric phase in double-pass signal could be of the same order of magnitude as the single-pass signal.\\
\indent In this simulation model, the distributions are assumed poloidally symmetric and toroidally homogeneous. However, the realistic plasma distributions are asymmetric due to the break of poloidal symmetry and different kinds of local fluctuations. In order to assess the reasonable range of the geometric phase, the simulation is carried out using single-pass setup. And typical high operation parameters in EAST tokamak are utilized.\\

\subsection{\label{section 4.1}Simulation results: Influence of detecting position, density, and wave frequency on the geometric phase}
\begin{figure}[htbp]
\includegraphics[width=.5\textwidth]{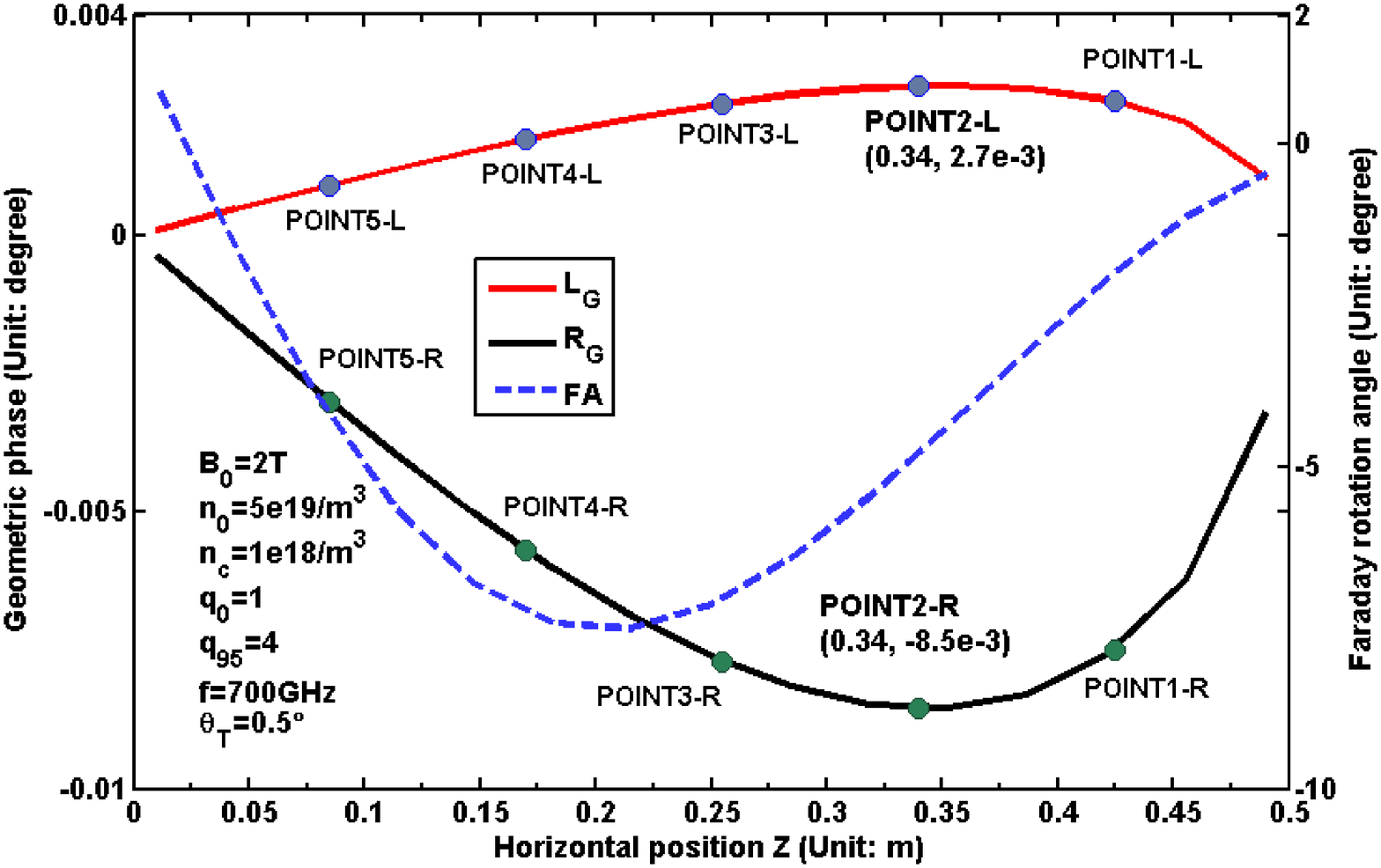}
\caption{The influence of detecting positions on the geometric phase and the Faraday rotation angle. The left ordinate together with the red curve labeled with $L_G$ depicts the geometric phase of L-wave. The left ordinate together with the black curve labeled with $R_G$ denotes the geometric phase of R-wave. The right ordinate together with the blue dashed curve labeled with $F_A$ denotes the Faraday rotation angle.}
\label{fig:2a}
\end{figure}

\begin{figure}[htbp]
\includegraphics[width=.5\textwidth]{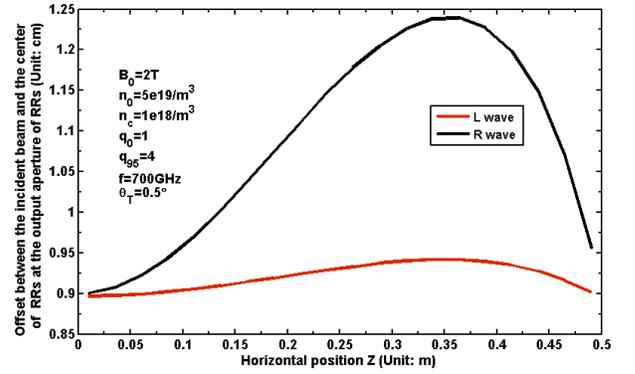}
\caption{The influence of detecting positions on the offset between the incident beam and the center of retro reflectors (RRs) at the output aperture of RRs.}
\label{fig:2b}
\end{figure}

\indent Because of the non-uniformity of the magnetized plasmas, signals from different channels of POINT system undergo different trajectories. Different dispersion relations and torsions along these trajectories lead to different magnitudes of the geometrical phases and the dynamical phases. According to the simulation results based on the optical design of the multi-channel POINT system, it is expected that the maximum geometric phase takes place in the second channel of POINT, see Fig.~\ref{fig:2a}. In Fig.~\ref{fig:2a}, it can be observed that the geometric phase is much larger for the R-wave than L-wave for some chords. This is because the torsion of the trajectory of the R-wave is much larger than that for the L-wave. It reflects the geometric origin of the geometric phase and its close dependence on the geometric property of the wave trajectory. At the output aperture of RRs, the offset between the incident R-wave and the center of RRs is different from the corresponding offset between the incident L-wave and the center of RRs. This difference demonstrates the offset between the R-wave and L-wave caused by refractive effect. As shown in Fig.~\ref{fig:2b}, the offset between the R-wave and L-wave at the output aperture of RRs is several millimeters under typical operation parameters of EAST tokamak. Therefore the offset between the R-wave and L-wave caused by refractive effect is of the order of millimeters in the overall propagation.

\begin{figure}[htbp]
\includegraphics[width=.47\textwidth]{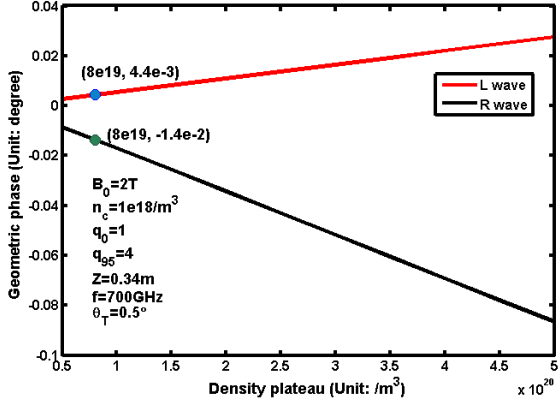}
\caption{Influence of density plateau on geometric phase.}
\label{fig:3}
\end{figure}

\begin{figure}[htbp]
\includegraphics[width=.47\textwidth]{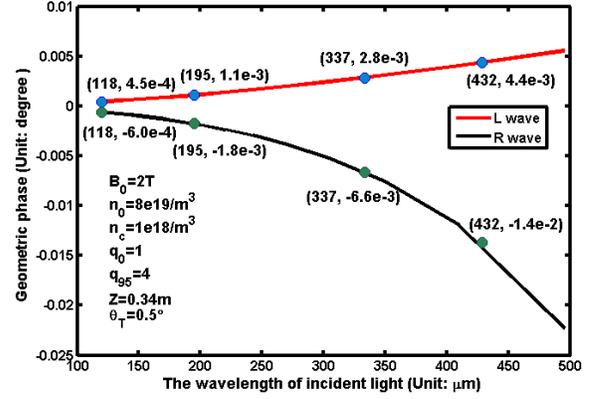}
\caption{Influence of wave frequency on geometric phase.}
\label{fig:4}
\end{figure}

\indent The torsion of wave trajectories becomes more evident with the increase of the density gradient. Therefore, the geometric phase looks larger with higher density plateau, see Fig.~\ref{fig:3}. The core density of EAST Tokamak can reach $8\times10^{19}/m^3$ at the present stage. With a quadratic density profile, that is $N_n\!=\!2$ in Eq.~\ref{eq:14}, the geometric phase in the second channel of POINT system is of the order of 0.01 degree, see Fig.~\ref{fig:3}. And the theoretical value of Faraday rotation angle in this channel is about 7.47 degrees under the same parameters.\\
\indent To evaluate the range of the geometric effect in various probe beams, the geometric phases are also calculated using different incident frequencies. The geometric phase decreases with the increase of the incident frequency, see Fig.~\ref{fig:4}. This is because the deviation of trajectories is weakened as increase of the wave frequency. According to the signals from waves with regular detection wavelengths, the geometric phases are in the range from $1\times10^{-3}$ degree to $1\times10^{-2}$ degree, see Fig.~\ref{fig:4}. This result is much smaller than the general estimation in Ref.~$7$. Besides the estimation parameters being different, this is mainly due to the trajectory-dependent property of the geometric phase. Although the geometric phase comes from the first-order term, the same as the dynamical phase of Faraday rotation angle, its coefficient still depends on the specific wave trajectory. The coefficient of the geometric phase term turns out to be very small in the case where the incident wave is perpendicular to the toroidal field. In the next subsection, it will show that for incident probe beams with larger toroidal angle the coefficient of the geometric phase term will increase and agree with the estimation in Ref.~$7$. We also note that the change of the wavelength of the incident waves can alter the magnitude of the geometric phase. The geometric phase does not depend on the wavelength directly through the dispersion relations like the dynamical phase. However, different wavelengths will lead to different wave trajectories, especially in complex plasma distributions and 3D configuration of magnetic fields. The geometric phases are thus affected through the wave trajectories indirectly by the wavelength.\\

\subsection{\label{section 4.2}Simulation results: The influence of incident angle on the geometric phase}
\begin{figure}[htbp]
\includegraphics[width=.5\textwidth]{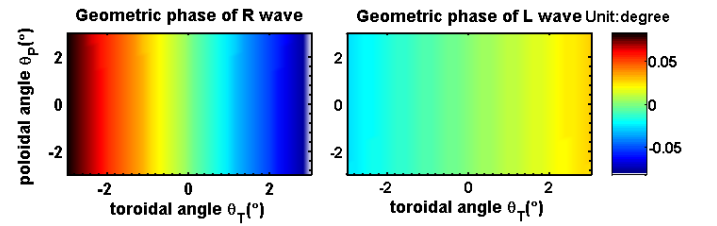}
\caption{Different effect of toroidal and poloidal angles on geometric phase.}
\label{fig:5}
\end{figure}

\begin{figure}[htbp]
\includegraphics[width=.47\textwidth]{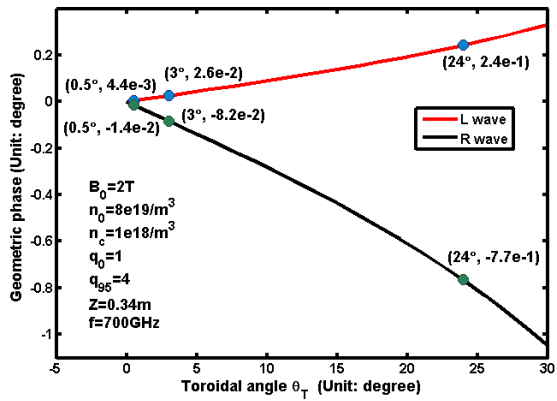}
\caption{The plot of geometric phases for L- and R- waves versus the toroidal angle.}
\label{fig:6}
\end{figure}

\indent The incident angle can be changed in both the poloidal and the toroidal planes. Their influences on the geometric phase can be evaluated separately. In toroidal coordinate system, the reflected wave vector by RRs can be expressed as
\begin{eqnarray}
\label{eq:15}
\left\{
\begin{aligned}
K_R&=kcos\theta_T\ast cos\theta_P, \\
K_Z&=kcos\theta_T\ast sin\theta_P, \\
K_\varphi&=ksin\theta_T,
\end{aligned}
\right.
\end{eqnarray}
\indent where $\theta_T$ is the angle between the incident beam and the poloidal plane, and $\theta_P$ is the angle between the incident beam and the toroidal plane. By substituting Eq.~\ref{eq:15} into the ray tracing equations, the influences of toroidal and poloidal angles on the geometric phase can be numerically calculated. The geometric phases of both R- and L- waves change significantly with the variances of the toroidal angle. But the change of the poloidal angle brings less impact on the geometric phase, see Fig.~\ref{fig:5}. So the geometric phase in POINT system can be greatly enhanced by manipulating the toroidal angle.\\
\indent In normal setup, the probe beam of poloidal Polarimeter-Interferometer system is perpendicular to the toroidal field to eliminate the toroidal field component along the wave trajectory. The realistic toroidal angle is less than 0.5 degree. The absolute value of geometric phase increases to 0.05 degree when the toroidal angle increases to 3 degree, and increases to 0.5 degree when the toroidal angle increases to 24 degree, see Fig.~\ref{fig:6}. Therefore, the geometric phase can be enhanced to a measurable level by increasing the toroidal angle.\\

\section{\label{section 5}Conclusions}
\indent According to our simulation, the maximum geometric phase in the present setup of EAST POINT system is of the order of 0.01 degree. Since the present resolution of EAST POINT system is about 0.1 degree\cite{Liu2014Faraday}, the verification of the geometric phase requires further improvement of the diagnostic accuracy. On the other hand, the geometric phase can be increased to a measurable level by increasing the toroidal angle in POINT system.\\
\indent In realistic situations, the signal processing methods become more complex with larger toroidal angle. When the angle between the incident beam and the poloidal plane becomes bigger, the dynamical phase of signal increases because the field component along the wave trajectory grows. The Cotton-Mouton and Faraday effects are no longer combined linearly when polarimetric effects are large.\cite{Segre1998Review} Consequently, the separation of Cotton-Mouton effect and Faraday effect, as well as the separation of the geometric phase and the dynamic phase effect, turns out to be more complex.\cite{Segre2006Derivation,Guenther2004Approximate}\\
\indent In the next step, some special designs can be applied to carry out the experimental measurement of the geometric phase. Firstly, the diagnostic resolution of POINT system will be enhanced. At the present stage, the effective ways to raise the diagnostic resolution includes the adjustment of the collinear paths of L- and R- waves outside the plasma, the elimination of stray light and crosstalk, precise calibration of the polarization state, and shielding of the mechanical vibration and electromagnetic interference. Using these methods, the POINT diagnostic resolution can be enhanced to around 0.05 degree. Then the geometric phase can easily increase to a measurable level by setting the toroidal angle to about 3 degree. Secondly, an optimum optical trajectory can be designed to enhance the geometric phase. For example, by increasing the propagation distance in plasma or the density gradient by inducing plasma transport barrier, the geometric phase can be effectively enhanced. The existing EAST POINT system offers an excellent hardware platform for the measurement of the geometric phase in plasma system.\\
\indent In the operation condition ($B_T$ = 5.5 T, core $n_e$ = 3$\times10^{20}/m^3$, edge $n_e=5\times10^{18}/m^3$, divertor $n_e$=5$\times10^{21}/m^3$, q=0.5-5, $\lambda$=118$\mu$m, $\theta_T$=0.5 degree) of the poloidal polarimeter system in ITER Tokamak \cite{Donne2004Poloidal,Donn2007Chapter}, the geometric phase in the chord via the equatorial port is of the order of 0.01 degree, and the geometric phase in the chord via the divertor region is of the order of 0.1 degree. Considering that accurately Faraday Effect measurement is necessary to reconstruct the safety factor profile with required accuracy \cite{Imazawa2011A}, the conscious controlling the geometric effect at a lower level is beneficial to acquire accurately q-profile measurement in the next generation tokamak devices.
\\

\begin{acknowledgments}
This work is supported by the National Magnetic Confinement Fusion Program of China with contract $No.2012GB101002$, $No.2014GB106002$, the National Nature Science Foundation of China with contract $No.11375237$, $No.11305171$, $No.11575185$, $No.11575186$, ITER-China Program with contract $No.2015GB111003$, $No.2014GB124005$.
\end{acknowledgments}
\bibliography{reference}

\end{document}